\documentclass[12pt]{article}

\def\be{\begin{equation}}
\def\ee{\end{equation}}
\def\ba{\begin{array}{c}}
\def\baa{\begin{array}{ll}}
\def\ea{\end{array}}

\def\ben{$$}
\def\een{$$}
 
\begin{document}

\titlepage

  \begin{center}{\Large \bf

Scattering theory using smeared non-Hermitian potentials

 }\end{center}

\vspace{5mm}

  \begin{center}

{\bf Miloslav Znojil}

 \vspace{3mm}

Nuclear Physics Institute ASCR, 250 68 \v{R}e\v{z}, Czech
Republic\footnote{ e-mail: znojil@ujf.cas.cz}\\

 \vspace{3mm}

\end{center}

\vspace{5mm}


\section*{Abstract}

Local non-Hermitian potentials $V(x)\neq V^*(x)$ can, sometimes,
generate stable bound states $\psi(x)$ at real energies.
Unfortunately, the idea [based on the use of a non-Dirac {\it ad
hoc} metric $\Theta(x,x') \neq \delta(x-x')$ in Hilbert space]
cannot directly be transferred to scattering due to the related loss
of the asymptotic observability of $x$ [cf. H. F. Jones, Phys. Rev.
D 78, 065032 (2008)]. We argue that for smeared (typically,
non-local or momentum-dependent) potentials $V \neq V^\dagger$ this
difficulty may be circumvented. A return to the usual (i.e., causal
and unitary) quantum scattering scenario is then illustrated via an
exactly solvable multiple-scattering example. In it, the anomalous
loss of observability of the coordinate remains restricted to a
small vicinity of the scattering centers.

 \vspace{9mm}%

\noindent PACS  03.65.Nk, 03.80.+r, 11.55.Ds;
%

\vspace{9mm}
%

\newpage

\section{Introduction 
}

An intuitive understanding of various physical aspects of scattering
can be facilitated when one turns attention to simplified,
one-dimensional schematic models of experimental setup. An extremely
exciting toy-model scenario has recently been proposed and analyzed
in ref.~\cite{Jonesdva}. One of the most elementary and popular
delta-function potentials $V_0(x) = -\alpha\,\delta(x)$ has
tentatively been combined with a remote non-Hermitian interaction.
The purpose of this {\it Gedankenexperiment} has been formulated as
a study of an ``interface" between Hermitian and non-Hermitian
components of the potentials exemplified by the superposition
 \be
 V(x) =V_0(x)+{\rm i}\beta\, [\delta(x-L) -
 \delta(x+L)]\,, \ \ \ \ \ L \gg 1.
 \label{jone}
 \ee
Our present text offers an immediate continuation of this project.
We feel motivated by the occurrence of many open questions in such a
setting. In particular, the results of ref.~\cite{Jonesdva}
indicated that it might be rather difficult to keep a non-Hermitian
interaction model short-ranged and compatible with the standard
requirements of a local and causal physical interpretation of
incoming and/or scattered waves.

Our present answer to these compatibility questions will be
predominantly affirmative. More precisely, we shall emphasize that
the number of problems which arose during the analysis of  potential
(\ref{jone}) may be attributed to its strict locality. We shall
propose and advocate the replacement of the strictly local
interaction operators $V\equiv V(x)$ by their slightly smeared
descendants tractable as weakly momentum-dependent operators. For
illustration purposes we shall use interactions $V\neq V^\dagger$
given by eq.~(\ref{propo}) in section \ref{jednotka} below.

The latter choice of amended model will preserve its maximal
similarity with the original potential of ref.~\cite{Jonesdva}.
Firstly, the role played by the real ``measure of non-Hermiticity"
$\beta$ in eq.~(\ref{jone}) will be transferred to another real
coupling constant $g$. Secondly, the distance $L$ between the
strictly localized interaction points of eq.~(\ref{jone}) will be
replaced by a variable integer ${\cal N}$ representing a separation
distance between two not entirely local, ``smeared" domains of
support of our interaction $V$.

We decided to parallel the majority of quantitative results of
ref.~\cite{Jonesdva} by their close and explicit analogues. The
major problems resulted from the manifest non-Hermiticity of the
interaction which implies, for eq.~(\ref{jone}) at least, the
necessity of a drastic change of the concept of the coordinate. In
ref.~\cite{Jones} this mathematical result has been identified as a
source of a deep conflict between the use of $x$ in eq.~(\ref{jone})
(i.e., in the ``input" definition of the interaction) and,
simultaneously, in the asymptotic boundary conditions for the
one-dimensional scattering,
 \be
 \psi(x) =
 \left \{
 \begin{array}{lr}
 e^{i\kappa x}+R\,e^{-i\kappa x}\,,\ \ \ \ \
 & x \ll -1\,,\\
 T\,e^{i\kappa x}\,,\ \ \ \ \ \ \ \ \ \ \
 &x \gg 1\,.
 \ea
 \right .
 \label{scatbc}
 \ee
It is necessary to keep in mind that the core of this conflict does
not lie in the formalism of quantum mechanics itself. Formally, no
problems occur since all the unitary transformations of a given
model (cf., e.g., a non-local free-motion example given in section 5
of ref.~\cite{cubic}) {\em must} lead to equivalent physical
predictions.

The differences in predictions can only occur when non-equivalent
definitions of the dynamics are being compared. This is precisely in
this sense that the difficulties emerged in
refs.~\cite{Jonesdva,Jones} where a simultaneous validity of  both
the definition (\ref{jone}) of the {\em local} physical interaction
$V\neq V^\dagger$  and   an {\it a priori} assignment (\ref{scatbc})
of the usual physical meaning to the {\em local} free waves $\exp
\pm {\rm i}\kappa x$ in asymptotic domain has been required.

A key to the resolution of this misunderstanding has been described
in our paper \cite{prd} where we showed that there exist
non-Hermitian models where the simultaneous validity of the
phenomenological postulates (\ref{jone}) and (\ref{scatbc}) {\em
can} be achieved after a certain modification of their respective
forms. In this sense, our present paper will just amend and
strengthen the argumentation of ref.~\cite{prd}. Indeed, in a way
emphasized by the note added in proof in \cite{Jonesdva}, our old
model ``still involved a departure from standard quantum mechanics
at large distances".

The present final resolution of the  conflict between the locality
of forces and waves will rely on a nontrivial extension of the class
of interactions accompanied by an enhancement of efficiency of
necessary mathematics. These technical details will be described in
sections \ref{secV} and \ref{secIV} and in two Appendices. {\it In
nuce}, the Runge-Kutta coordinate-discretization method \cite{RK}
will be shown superior, for the given purpose, to the usual
perturbation expansions as employed, e.g., in ref.~\cite{Jonesdva}.
In the latter study of eq.~(\ref{jone}), for this reason, {\em both}
the inverse length $1/L$ {\em and} the variable coupling constant
$\beta$ had to be assumed small. In contrast, the variability of
both our present parameters $g$ and ${\cal N}$ will be, within their
respective physical ranges, unrestricted.

The presentation of our explicit scattering solutions in section
\ref{secIV} will confirm the full consistency and unitarity of the
scattering in our amended class of non-Hermitian models. In summary
(section \ref{summary}) several comments will finally be added
clarifying the proposed changes of theoretical perspective in a
broader, less model-dependent context.


\section{Towards the short-ranged non-Hermiticities\label{jednotka}}

\subsection{Runge-Kutta discretization}

We may treat any one-dimensional Schr\"{o}dinger equation
 \be
 -\frac{d^2}{dx^2}\,\psi(x)+
  V(x)\,\psi(x) =E\,\psi(x)\,,
 \ \ \ \ \ \ \ x \in (-\infty,\infty)
 \label{SEloc}
 \ee
with a local and real or complex potential $V(x)$ as a continuous
$h\to 0$ limit of its difference-equation approximation defined
along the Runge-Kutta doubly infinite lattice of discrete
coordinates $x=x_k=k\,h$, $ k = 0, \pm 1, \ldots$,
 \be
 -\frac{\psi(x_{k-1})-2\,\psi(x_k)+\psi(x_{k+1})}{h^2}+V(x_k)\,
 \psi(x_k)
 =E\,\psi(x_k)\,.
 \label{SEdis}
 \ee
Approximate wave functions may be then constructed via the methods
of linear algebra re-parametrizing, incidentally, the real energies
$E= (2 -2 \cos \varphi )/h^2$ in terms of a real angle
$\varphi=\varphi(E)\,\in (0,\pi)$. The scattering boundary
conditions (\ref{scatbc}) may and should be re-written in their
discrete version,
 \be
 \psi(x_m) =
 \left \{
 \begin{array}{ll}
 e^{i\,m\,\varphi}+R\,e^{-i\,m\,\varphi}\,,\ \ \ \ \
 \ \ \ \ \ \
 & m \leq -M\ll -1\,,\\
 T\,e^{i\,m\,\varphi}\,,\ \ \ \ \ &
 m \geq M-1\,.
 \ea
 \right .
 \label{discatbc}
 \ee
One could easily discretize the ultralocal non-Hermitian toy model
(\ref{jone}) and confirm the discouraging conclusions, formulated in
ref.~\cite{Jonesdva}, that one can ``no longer talk in terms of
reflection and transmission coefficients" so that ``the only
satisfactory resolution [of dilemmas] is to treat the non-Hermitian
scattering potential as an effective one, and work in the standard
framework of quantum mechanics, accepting that this effective
potential may well involve the loss of unitarity" \cite{Jonesdva}.

The loss of unitarity need not necessarily be perceived as a
weakness of the theory, especially when one deals ``with a subsystem
of a larger system whose physics has not been taken fully into
account" \cite{Jonesdva}. In this sense one may perceive
eq.~(\ref{jone}) with a local non-Hermitian interaction $V(x) \neq
V^*(x)$  as an ``effective theory", i.e., as an {\em incomplete}
picture of physical reality. This philosophy finds interesting
phenomenological applications ranging from classical optics
\cite{Christo} or models with supersymmetry \cite{Siegl} up to the
manifestly non-unitary scattering models in quantum phenomenology
\cite{Cannata} and up to the descriptions of open systems in nuclear
and solid state  physics \cite{Rotter} and in quantum
cosmology~\cite{Kamen}. Nevertheless, in an alternative,
theoretically much more ambitious approach to the localized
non-Hermitian scattering potentials one should insist on the
conservation of a suitable current and, hence, on the strict
unitarity of the scattering realized by the asymptotically
observable free plane waves.

\subsection{Nonlocal updates of potentials}

The first (and the least sophisticated) quantitative non-Hermitian
model satisfying the above requirements has been constructed in our
paper \cite{prd}. We replaced equation~(\ref{SEdis}) by its
generalization where the interaction operator $V$ acquired the
nearest-neighbor form,
 \ben
  -
 \frac{\psi(x_{k+1})-2\,\psi(x_{k})+\psi(x_{k-1})}{h^2}+
  \ \ \ \ \ \ \ \ \ \ \ \
 \ \ \ \ \ \ \ \ \ \ \ \
 \een
 \be
  \ \ \ \ +V_{k,{k+1}}\,\psi(x_{k+1})
 +V_{k,{k}}\,\psi(x_{k})
 +V_{k,{k-1}}
 \,\psi(x_{k-1})
 =E\,\psi(x_k)\,.
 \label{diskretni}
 \ee
After a re-scaling of the Hamiltonian $H=-d^2/dx^2+{V}$ by an
inessential numerical factor $h^2$ we obtained
 \be
 H= -\triangle
 + V\,,\ \ \ \
 -\triangle=
 \begin{array}{|cccccc|}
 \hline
 \ddots &\ddots &&&& \\
 \ddots &2&-1&&&\\
 &-1&2&-1&&\\
 &&-1&2&-1&\\
 &&&-1&2&\ddots \\
  &&&&\ddots & \ddots \\
 \hline
 \ea\,
 \label{dvanact}
 \ee
and choose the following, minimally non-local  potential
 \be
V= V^{(a,b,c,\ldots )}=\left [\begin {array}{cccc|cccc}
 \hline
 &\ddots&{}&{}&{}&{}&{}&{}{}{}{}\\
 \ddots&{}& -c &{}&{}&{}&{}&{}{}{}{}\\
 &c&{}&-b&{}&{}&{}&{}{}{} \\
 &{}&b&{}&-a&{}&{}&{}{}{}{}\\
 \hline
 {}&{}&{}&a&{}&-b&{}&{}{} \\
 {}&{}&{}&{}&b&{}&-c&{}{}{}\\
 {}  &{}&{}&{}&{}&c&{}&\ddots{}{}\\
 {}&{}&{}&{}&{}&{}&\ddots&{}\\
 \hline
 \end {array}\right ]\,.
 \label{muli}
 \ee
The resulting multiparametric Hamiltonian $H=-\triangle+
V^{(a,b,c,\ldots )}\neq H^\dagger$ remains manifestly non-Hermitian
in the ``friendly" Hilbert space ${\cal H}^{(F)}$ endowed with the
usual inner product
 \be
 \langle \psi|\phi \rangle^{(F)}=
 \sum_k\,\psi^*(x_k) \phi(x_k)=\langle \psi|\phi
 \rangle
 \,\ \ \ \ \ \ \ {\rm in}
 \ \  {\cal H}^{(F)}\,.
 \label{diracs}
 \ee
Note that the summation would only be replaced by the integration in
the continuous limit $h\to 0$. Now, the key point is that the {\em
same} operator $H$ may be found Hermitian after one moves into {\em
another} Hilbert space ${\cal H}^{(S)}$. In the latter space the
definition of the inner product must be different and more general,
 \be
 \langle \psi|\phi \rangle^{(S)}=
 \sum_k\,\sum_n\,\psi^*(x_k) \Theta_{k,n}\,\phi(x_n)=
 \langle \psi|\Theta|\phi \rangle:=\langle\!\langle \psi|\phi
 \rangle
 \ \ \ \ {\rm in}
 \ \  {\cal H}^{(S)}\,.
 \label{e777}
 \ee
The ``non-Dirac metric" matrix $\Theta=\Theta^\dagger$ must only
remain positive definite and compatible with the Hamiltonian in
question \cite{Geyer},
 \be
 H^\dagger \,\Theta = \Theta\,H
 \,.
 \label{quasiher}
 \ee
In the notation of ref.~\cite{SIGMA} one writes $H=H^\ddagger$
and speaks about a ``quasi-Hermiticity" \cite{Geyer} or
``pseudo-Hermiticity" \cite{Ali} or ``cryptohermiticity"
\cite{Smilga} of the Hamiltonian. In this context the core of the
message delivered by our paper \cite{prd} was that there exists a
metric-operator matrix $\Theta^{(a,b,c,\ldots )}$ which remains
compatible with our interaction model (\ref{muli}) {\em as well as}
with the asymptotic observability of  Runge-Kutta coordinates $x_k$.
This matrix has the following compact and fully diagonal form,
 \be
 \Theta^{(a,b,c,\ldots )}=\left [\begin {array}{cccc|cccc}
 \hline
 \ddots&&{}&{}&{}&{}&{}&{}{}{}{}\\
 &{\theta_{-5}}&  &{}&{}&{}&{}&{}{}{}{}\\
 &&{\theta_{-3}}&&{}&{}&{}&{}{}{} \\
 &{}&&{\theta_{-1}}&&{}&{}&{}{}{}{}\\
 \hline
 {}&{}&{}&&{\theta_1}&&{}&{}{} \\
 {}&{}&{}&{}&&{\theta_3}&&{}{}{}\\
 {}  &{}&{}&{}&{}&&{\theta_5}&{}{}\\
 {}&{}&{}&{}&{}&{}&&\ddots\\
 \hline
 \end {array}\right ]\,.
 \label{formul}
 \ee
Its elements are given by closed formulae,
 \ben
 \theta_{\pm 1}=(1\pm a)(1-b^2)(1-c^2)(1-d^2) \ldots\,,
 \een
 \ben
 \theta_{\pm 3}=(1\pm a)(1\pm b)^2(1-c^2)(1-d^2) \ldots\,,
 \een
 \ben
 \theta_{\pm 5}=(1\pm a)(1\pm b)^2(1\pm c)^2(1-d^2) \ldots\,.
 \een
One arrives at a causality-observing physical picture of scattering
based on a clear separation of the ``in" and ``out" solutions not
only in Hilbert space ${\cal H}^{(F)}$ but also in Hilbert space
${\cal H}^{(S)}$.

In our subsequent paper \cite{FT} the next step has been made. In
the spirt of eq.~(\ref{jone}) we simulated the existence of several
separate point interactions. Unfortunately, the construction of the
metric only remained feasible under a very specific, left-right
symmetric arrangement of the set of interaction centers. Sometimes,
this type of symmetry is  being called ${\cal PT}-$symmetry, for
reasons and with motivations which are thoroughly explained
elsewhere \cite{Carl}.

Our present continuation of development of the multiple-scattering
idea will be based on a return to asymmetric models, allowing an
independence of arrangement of {\em several} spatially separated
scatterers.  Paradoxically, the transition to asymmetric
realizations of the set of interaction centers will be accompanied
by a simplification of analysis of their mutual interference.

\subsection{Limiting transition
to continuous coordinates $h\to 0$ }

For a quantitative specification of the extent of  nonlocality
induced by multiparametric matrices $V^{(a,b,c,\ldots )}$ of
eq.~(\ref{muli}) let us start from the simplest,
coordinate-independent model where $a \approx b \approx c \approx
\ldots$. Then, the limiting transition to $h=0$ converts operator
$V^{(a,a,a,\ldots )}$ into the first power of the momentum,
$V^{(a,a,a,\ldots )} \sim d/dx$. In the subsequent step one may
re-introduce a weak coordinate-dependence (with $ a \neq b \neq
\ldots$) and evaluate the continuous limit perturbatively. Locally,
the limit $h \to 0$ will preserve the same leading-order approximate
proportionality of the coordinate-dependent potential to the
momentum.

Admitting an unconstrained variability of the parameters in matrices
$V^{(a,b,c,\ldots )}$ we obtain some less trivial coordinate- and
momentum-dependent operators. For the sake of brevity let us
restrict similar considerations solely to the models with just a few
nonvanishing coupling parameters. Then, the limiting transition $h
\to 0$ will certainly lead to point interactions. Their explicit
definition will be given precisely by the matching of the wave
functions. Just a slightly more complicated alternative to the
delta-function point-interaction model (\ref{jone}) of
ref.~\cite{Jonesdva} will be obtained. Our Appendices A and B may be
consulted for illustration of some technical aspects of such a type
of matching recipe.

Our illustrative toy potential (\ref{muli}) has not been too well
designed for phenomenological purposes since it did not allow us to
remove the spatial asymmetry from the related metric matrix
~(\ref{formul}),
 \be
 \frac{\theta_{-k}}{\theta_{k}}
 =\frac{(1- a)(1- b)^2(1- c)^2 \ldots}
 {(1+ a)(1+ b)^2(1+ c)^2 \ldots}\,.
 \label{clo}
 \ee
The effect of the localized non-Hermiticity in $H$ remained
long-ranged.

\section{Toy model \label{secV}}

Equation (\ref{clo}) indicates that the flow of the probability is
different to the left and to the right of the scattering center. A
weaker form of this shortcoming characterizes also the ${\cal
PT}-$symmetric models of ref.~\cite{FT} where the metric remained
rescaled (i.e., non-Dirac, $\Theta_{k,k}\neq 1$) along the spatial
interval(s) separating the individual scatterers. This encouraged us
to perform a series of computer-assisted trial-and-error experiments
leading, at the end, to our present interaction-matrix candidate
 \be
 V^{(g,{\cal N})}=
 \left [\begin {array}{cc|ccc|cc|ccc|cc}
 \hline
 &\ddots&{}&{}&{}&{}&{}&{}&{}&{}{}&{}&{}\\
 \ddots& & {0} & & & & & & &  & & \\
 \hline
 &{0}& &-g& & & & & & & &  \\
 & &g& &g& & & & &  & & \\
 && &-g& &{0}& & & & &  & \\
 \hline
 & & &  &{0}& &\ddots& & & &  & \\
 & & &  &&{\ddots}&&{{0}}& & &  & \\
 \hline
 & & & & & &{0}& &-g& &   & \\
 & & & & & \multicolumn{2}{c|}{ \underbrace{\ \ \ \ \ \ \ \  \ \ \ \ \ \ } }
 &g& &g&   &\\
 & & & & &
 \multicolumn{2}{c|}{ \rm large\ gap, }
  & &-g& &{0}&   \\
 \hline
 & & & &   &
 \multicolumn{2}{c|}{   2{\cal N}+1 }
 & & &{0}& &\ddots  \\
 & & & &   &
 \multicolumn{2}{c|}{    {\rm \ columns }}
 & & & & \ddots&  \\
 \hline
 \end {array}\right ]\,
 \label{propo}
 \ee
where each scatterer is  simulated by three-dimensional submatrix.
Although our particular model (\ref{propo}) comprises just two
localized interaction centers at $x_{\pm ({\cal N}+2)}$, we shall
not consider three (like in eq.~(\ref{jone})) or more individual
scatterers because such a generalization would remain routine, not
necessitating any significant further technical improvements of our
method.

\subsection{Metric $\Theta^{(g,{\cal N})}$ with localized anomalies}

Using heuristic arguments we arrived at ansatz (\ref{propo}) and
studied the scattering solutions. Schr\"{o}dinger equation with the
smallest gaps ${\cal N}$ has been studied first of all. A sample of
these calculations may be found collected in Appendices A and B
below.  They demonstrate that one of specific merits of
eq.~(\ref{propo}) lies in a maximal simplicity of necessary
algebraic manipulations.

The second merit of the choice of eq.~(\ref{propo}) can be seen in
its generic character. One can add several further interaction
submatrices of the same form without worsening the feasibility of
the calculations. On this background, without any real loss of
generality we restricted our attention just to the first nontrivial
example which is characterized by the occurrence of the mere two
remote centers of interaction.

We decided to construct all the eligible metric matrices as
linear-algebraic solutions of eq.~(\ref{quasiher}). After we imposed
the condition of the compatibility of $\Theta$ with the asymptotic
observability of the coordinate we revealed that our present models
$V^{(g,{\cal N})}$ can be assigned the diagonal metric operators of
the same doubly infinite diagonal matrix form
 \be
 \Theta^{(g,{\cal N})}=
 \left [\begin {array}{cc|ccc|cr|ccc|cc}
 \hline
 \ddots&&{}&{}&{}&{}&{}&{}&{}&{}{}&{}&{}\\
 & 1&  & & & & & & &  & & \\
 \hline
 && 1&& & & & & & & &  \\
 & &&\frac{1+g}{1-g} && & & & &  & & \\
 && && 1&& & & & &  & \\
 \hline
 & & &  && 1&& & & &  & \\
 & & &  &&&{\ddots}\ \ \ \ \ \ \ && & &  & \\
 & & &  &&&\ \ \ 1&& & &  & \\
 \hline
 & & & & & && 1&& &   & \\
 & & & & & & &&\frac{1+g}{1-g} &&   &\\
 & & & & &
 \multicolumn{2}{c|}{ \underbrace{\ \ \ \ \ \ \ \  \ \ \ \ \ \ } }
  & & & 1&&   \\
 \hline
 & & & &   &
 \multicolumn{2}{c|}{  = 2{\cal N}+1 }
 & & && 1&  \\
 & & & & &
 \multicolumn{2}{c|}{{\rm (``distance") }} & & & &&\ddots \\
 \hline
 \end {array}\right ]\,.
 \label{theme}
 \ee
This metric differs from the Dirac's $\Theta^{(Dirac)}=I$ {\em
solely } at the centers of the non-vanishing three-by-three
submatrices simulating the non-Hermitian point-like scatterers.

\subsection{Single-center limit $h\to 0$ at ${\cal N}=-1$}

The picture of the scattering as offered by our toy
potential~(\ref{propo}) and by the related metric matrix
(\ref{theme}) depends on the Runge Kutta discretization length
$h>0$. Once we demand that a measured distance between two
scattering centers is a macroscopic constant $L$, our parameter
${\cal N}$ must grow with the decrease of $h$ as $L/h$. {\it Vice
versa}, the use of an $h-$independent ${\cal N}$ will only lead to a
single-centered scatterer. In the latter scenario the scattering is
realized by a ``quasi-local" potential. Its explicit specification
will depend on the $h-$dependence of ${\cal N}= {\cal N}(h)$. It
remains compatible with $L=0$ whenever $h\,{\cal N}(h) \to 0$ for $h
\to 0$. Such a flexibility may make the interactions better suited
for fine-tuning, say, of the strength of non-localities and/or of
the extent of the violation of conservation laws at short distances,
etc.

For illustration purposes let us pick up the elementary example of
Appendix A. Relaxing the specification of some concrete asymptotic
boundary conditions  let us re-interpret its ``distant" wavefunction
components as an arbitrary free wave $\psi(x)$. At $x\leq x_{-2}$ or
$x\geq x_2$ this yields the coincidence of symbols
 \ben
 U_{-m} \ = \  \psi(x_{-m}):=\psi_{-m}^{(free)}\,,
 \ \ \ \ \ \ \
 L_{m}  \ = \  \psi(x_m):=\psi_{+m}^{(free)}\,,
 \ \ \ \ \  m \geq {\cal N}+3=2\,
 \een
respectively. Next, the first and last matching condition  extend
both the latter assignments by one more step,
 \ben
 U_{-1} \ = \  (1+{g})^{}\,\psi(x_{-1})\,,\ \ \ \ \ \ \ \
 L_1 \ = \  (1+{g})^{}\,\psi(x_{+1})\,.
 \een
Finally, with $ \psi_0\ = \ \psi(x_0)$ we arrive at the three
dynamically nontrivial requirements
 \be
\left [\begin {array}{ccc}
     2\cos \varphi&-1&0
  \\
   -1+{g}^2&2\cos \varphi&-1+{g}^2
 \\
 0&-1&2\cos \varphi
  \end {array}\right ]\,
  \left [\begin {array}{c}
  U_{-1}
 \\
 (1-{g}^2)\,
 \psi_0
 \\
  L_1
 \end {array}\right ]=(1-{g}^2)\,
 \left [\begin {array}{c}
   \psi_{-2}^{(free)}
 \\0
 \\
   \psi_2^{(free)}
 \end {array}\right ]
 \,
 \label{urge}
 \ee
which define our wave function, implicitly, near the origin.
Tentatively, we may Taylor-expand
 \ben
 \psi_{-2}^{(free)}=\psi-2h\psi'+2h^2\psi''+\ldots\,,
 \ \ \
 (1+g)^{-1}U_{-1}=\psi-h\psi'+h^2\psi''/2+\ldots\,,
 \een
 \ben
 \ \ \
 \psi_0=\psi\,,
 \ \ \
 (1+g)^{-1}L_{1}=\psi+h\psi'+h^2\psi''/2+\ldots\,,
 \ \ \
 \psi_{2}^{(free)}=\psi+2h\psi'+2h^2\psi''+\ldots\,
 \een
and insert these approximants in eq.~(\ref{urge}), yielding
 \ben
 -(1-g)\,(\psi-2h\psi'+2h^2\psi'')+
 2\cos \varphi\,(\psi-h\psi'+h^2\psi''/2)
  -(1-g)\,
 \psi={\cal O}(h^3)\,,
 \een
 \ben
 -(1+g)\,(\psi-h\psi'+h^2\psi''/2)+
 2\cos \varphi\,\psi
  -(1+g)\,(\psi+h\psi'+h^2\psi''/2)
 ={\cal O}(h^3)\,,
 \een
 \ben
 -(1-g)\,(\psi+2h\psi'+2h^2\psi'')+
 2\cos \varphi\,(\psi+h\psi'+h^2\psi''/2)
  -(1-g)\,
 \psi={\cal O}(h^3)\,.
 \een
According to these relations, Schr\"{o}dinger equation
$V-E=\psi''/\psi$ would make quantity $V$ large and positive when
extracted from the combination of the first and third equation, or
large and negative when extracted from the middle equation. This
means that our tentative assumption about the smoothness of wave
functions near the origin leads to mathematical contradictions and
must be abandoned.

Let us now modify our assumptions, distinguish between the left and
right wave functions and set $A(x)=\psi(x-h)$ and $B(x)=\psi(x+h)$,
i.e.,
 \ben
 (1+g)^{-1}U_{-1}=A(0):=A\,,
 \ \ \
 (1+g)^{-1}L_{1}=B(0):=B\,.
 \een
Naturally,
 \ben
 \psi_{-2}^{(free)}\approx A-hA'+{\cal O}(h^2)\,,
 \ \ \
 \psi_{2}^{(free)}\approx B+hB'+{\cal O}(h^2)\,
 \een
while quantity $\psi_0$ acquires the two alternative first-order
representations,
 \ben
  \psi_{0}\approx A+hA'\approx B-hB'\,.
 \een
In the limit $h\to 0$ the latter relation yields
 \ben
 A=B\,,\ \ \ \ \ \ \ \ A'=-B'\,.
 \een
The insertion of our amended ansatzs in eq.~(\ref{urge}) leads just
to the three alternative versions of the requirement of smallness of
$A={\cal O}(h^2)$, $B={\cal O}(h^2)$ as well as of $\psi_0={\cal
O}(h)$.  Thus, in the continuous-coordinate extreme our simplest
${\cal N}=-1$ example degenerates to the opaque-wall-barrier
dynamics generated by an additional Dirichlet boundary condition
$\psi(0)=0$.

We see that the role of non-Hermiticity is, in our model with ${\cal
N}=-1$ at least, truly non-perturbative and dynamically highly
influential. This conclusion may independently be confirmed by the
inspection of the ${\cal N}=-1$ reflection and transmission
coefficients given in Appendix A. We believe that also beyond this
concrete example, at least some of its features will survive a
transition to more-center models and/or to the two-center models at
large separation distances ${\cal N}(h) = {\cal O}(1/h)$.

\section{The unitarity of the scattering at any ${\cal N}$ \label{secIV} }


The first encouraging surprise encountered during the inspection of
the discretized metric (\ref{theme}) is that it remains
asymptotically diagonal in the coordinate representation. This means
that the asymptotic coordinate $x$ remains observable. Moreover, the
range of influence of individual non-Hermitian scatterers is
shortened via eqs.~(\ref{propo}) and (\ref{theme}). Thus, the only
missing component of the whole picture are formulae for the
reflection and transmission coefficients, the determination of which
may start from the linear Schr\"{o}dinger equation for discretized
wave functions $\psi=\psi(x_k)=\psi_k$,
 \be
 H\,\psi=E\,\psi\,.
 \label{SE}
 \ee
In its light the validity of boundary conditions (\ref{discatbc})
can be prolonged to all the subasymptotic free-motion domain,
 \be
  \left . \begin{array}{l}
 \psi_{-m}= e^{-{\rm i}m\varphi}+ R\,e^{{\rm
 i}m\varphi}\,\equiv\,U_{-m},\\
 \psi_{+m}=T\,e^{{\rm i}m\varphi }
 \,\equiv\,L_{m}\,,
 \ea
 \right \}\ \ \
 m \geq {\cal N}+3\,.
 \label{discatbcrev}
 \ee
In parallel, for larger integers ${\cal N}={\cal N}(h)$ we may
profit from adding another free-motion ansatz at the  smaller
subscripts,
 \be
 \psi_k=C\,e^{{\rm i}k\varphi }+ D\,e^{-{\rm i}k\varphi}\,,\ \ \ \ \
 \ \ |k|\leq {\cal N}\,.
 \label{interna}
 \ee
One should add that the study of the large distances ${\cal N}\gg 1$
might be well motivated by its potential relevance in physics. In
particular, its feasibility could offer a guide for  simulation of
macroscopic non-localities, the presence of which could, in its
turn, lead to the violation of causality at  small distances. In
parallel it is important that the effect of our non-Hermitian $V$
can be kept localized. This means that in contrast to virtually all
of the published older models the simplicity of interaction
(\ref{propo}) enables us to return to the ``old-fashioned"
definitions of the reflection coefficient $R$ and transmission
coefficient $T$.

\subsection{The elimination of ${\cal N}$ from matching conditions.}

The second surprise offered by our example is that the matching
remains easy  even for remote interactions with ${\cal N}\gg 1$. In
order to show this, let us now assume that the distance $2\,{\cal
N}+1$ between two three-dimensional interaction submatrices in
(\ref{propo}) is arbitrary. We may abbreviate, in partitioned
notation,
 \ben
 V^{(g,{\cal N})}=
\left [\begin {array}{rrr|c|rrr} \hline
 0&g&0&\vec{0}^{\,T}&0&0&0
  \\ \!-g&0&\!-g&\vec{0}^{\,T}&0&0&0
 \\
 0&g&0&\vec{0}^{\,T}&0&0&0
  \\
  \hline
 \vec{0}&\vec{0}&  \vec{0}&\widehat{0}&\vec{0}&\vec{0}&\vec{0}
 \\
 \hline
 0&0&0&\vec{0}^{\,T}&0&g&0
 \\
 0&0&0&\vec{0}^{\,T}&\!-g&0&\!-g\\
 0&0&0&\vec{0}^{\,T}&0&g&0
 \\
 \hline
 \end {array}\right ]\,.
 \een
where $\widehat{0}$ denotes a null matrix (of dimension $2{\cal
N}+1$) and where $\vec{0}$ are null column vectors. The superscripts
$^{\,T}$ denote transpositions (i.e., row real vectors). In such a
notation one has to consider the following $2{\cal N}+7$ matching
conditions
 \ben
 {M}^{[{\cal N}]}( \varphi)\,
   \left [\begin {array}{c}
  \hline
  U_{-{\cal N}-3}\\
  U_{-{\cal N}-2}
  +
 \chi_{-2} \\
  U_{-{\cal N}-1}
 +
 \chi_{-1}
 \\
 \hline
 \overrightarrow{\psi}_0
 \\
 \hline
  L_{{\cal N}+1}+\chi_1 \\
  L_{{\cal N}+2}+
 \chi_{2} \\
  L_{{\cal N}+3}\\
  \hline
 \end {array}\right ]=
  \left [\begin {array}{c}
  \hline
  U_{-{\cal N}-4}\\
  0\\
   0\\ \hline
   \vec{0}\\ \hline 0\\
   0\\
  L_{{\cal N}+4}\\
  \hline
 \end {array}\right ]
 \,
 \een
where
 \ben
 M^{[{\cal N}]}( \varphi)=
\left [\begin {array}{ccc|c|ccc} \hline
    2\cos \varphi&-1-{g}&0&\vec{0}^{\,T}&0&0&0
 \\
    -1+{g}&2\cos \varphi&-1+{g}&\vec{0}^{\,T}&0&0&0
 \\
 0&-1-{g}&2\cos \varphi&\vec{a}^{\,T}&0&0&0
  \\ \hline
 \vec{0}&\vec{0}&  \vec{a}&\widehat{F}^{[{\cal N}]}&\vec{b}&\vec{0}&\vec{0}
 \\
  \hline
 {0}&{0}&0&\vec{b}^{\,T}&2\cos \varphi&-1-{g}&0
 \\
 0&0&0&\vec{0}^{\,T}&-1+{g}&2\cos \varphi&-1+{g}\\
 0&0&0&\vec{0}^{\,T}&0&-1-{g}&2\cos \varphi\\
 \hline
 \end {array}\right ]
 \,
 \een
and where $\vec{a}^{\,T}=(-1,0,\ldots,0)$ and
$\vec{b}^{\,T}=(0,\ldots,0,-1)$ are two $(2{\cal N}+1)-$dimensional
auxiliary row vectors. The other auxiliary ``free-motion" submatrix
$\widehat{F}^{[{\cal N}]}$ is tridiagonal and $(2{\cal
N}+1)-$dimensional. Its elements $2\cos \varphi$ along the main
diagonal are complemented by the elements $-1$ which lie along its
two neighboring diagonals.

\subsection{Exact solvability.}

What remains for us to demonstrate is that our model conserves the
global or asymptotic flow of probability, i.e., that one obtains
$|R|^2+|T|^2=1$ in spite of the manifest non-Hermiticity of the
Hamiltonian $H$. In this setting the final surprise comes with the
observation that the reflection and transmission coefficients are
obtainable in closed form. Even when the ``distance" parameter
${\cal N}$ is arbitrarily large, the use of ansatz (\ref{interna})
reduces the original set of $2{\cal N}+7$ matching conditions to the
following two independent matching conditions consisting of four
items each,
 \ben
\left [\begin {array}{cccc} \hline
    2\cos \varphi&-1-{g}&0&0
 \\
    -1+{g}&2\cos \varphi&-1+{g}&0
 \\
 0&-1-{g}&2\cos \varphi&-1
  \\
 0&0&-1&2\cos \varphi
 \\
 \hline
 \end {array}\right ]
 \,
   \left [\begin {array}{c}
  \hline
  U_{-{\cal N}-3}\\
  U_{-{\cal N}-2}
  +
 \chi_{-2} \\
  U_{-{\cal N}-1}
 +
 \chi_{-1}
 \\
 \psi_{-{\cal N}}\\
 \hline
 \end {array}\right ]=
  \left [\begin {array}{c}
  \hline
  U_{-{\cal N}-4}\\
  0\\
   0\\
   \psi_{-{\cal N}+1}\\
  \hline
 \end {array}\right ]
 \,,
 \een
 \ben
 \left [\begin {array}{cccc} \hline
    2\cos \varphi&-1&0&0
 \\
 -1&2\cos \varphi&-1-{g}&0
 \\
 0&-1+{g}&2\cos \varphi&-1+{g}\\
 0&0&-1-{g}&2\cos \varphi\\
 \hline
 \end {array}\right ]\,
   \left [\begin {array}{c}
  \hline
   {\psi}_{\cal N}
 \\
  L_{{\cal N}+1}+\chi_1 \\
  L_{{\cal N}+2}+
 \chi_{2} \\
  L_{{\cal N}+3}\\
  \hline
 \end {array}\right ]=
  \left [\begin {array}{c}
  \hline
  {\psi}_{{\cal N}-1}\\
  0\\
   0\\
  L_{{\cal N}+4}\\
  \hline
 \end {array}\right ]
 \,.
 \een
Out of this octuplet of equations, the first and last lines can be
solved,
 \ben
 (1+{g})\chi_{-2}=-{g}U_{-{\cal N}-2}\,,
 \ \ \ \ \
 (1+{g})\chi_{2}=-{g}L_{{\cal N}+2}\,.
 \een
This leads to the following two triplets of conditions
 \ben \left
 [\begin {array}{ccc} \hline
       2\cos \varphi&-1+{g}^2&0
 \\
 -1&2\cos \varphi&-1
  \\
 0&-1&2\cos \varphi
 \\
 \hline
 \end {array}\right ]
 \,
   \left [\begin {array}{c}
  \hline
  U_{-{\cal N}-2}
   \\
  U_{-{\cal N}-1}
 +
 \chi_{-1}
 \\
 \psi_{-{\cal N}}\\
 \hline
 \end {array}\right ]=
  \left [\begin {array}{c}
  \hline
  (1-{g}^2)
  U_{-{\cal N}-3}\\
   0\\
   \psi_{-{\cal N}+1}\\
  \hline
 \end {array}\right ]
 \,,
 \een
 \ben
 \left [\begin {array}{ccc} \hline
    2\cos \varphi&-1&0
 \\
 -1&2\cos \varphi&-1
 \\
 0&-1+{g}^2&2\cos \varphi\\
 \hline
 \end {array}\right ]\,
   \left [\begin {array}{c}
  \hline
   {\psi}_{\cal N}
 \\
  L_{{\cal N}+1}+\chi_1 \\
  L_{{\cal N}+2} \\
  \hline
 \end {array}\right ]=
  \left [\begin {array}{c}
  \hline
  {\psi}_{{\cal N}-1}\\
  0\\
   (1-{g}^2)
  L_{{\cal N}+3}\\
  \hline
 \end {array}\right ]
 \,.
 \een
Using the first and last equation we eliminate
 \ben
 (1-{g}^2)\,\chi_{-1}={g}^2\,U_{-{\cal N}-1}+{g}^2U_{-{\cal N}-3}\,,\ \
 \ \ \ \ \
 (1-{g}^2)\,\chi_{1}={g}^2\,L_{{\cal N}+1}+{g}^2L_{{\cal N}+3}\,.
 \een
The net result of these manipulations are the four relations
 \ben
 \left
 [\begin {array}{cc} \hline
 2\cos \varphi&-1
  \\
 -1&2\cos \varphi
 \\
 \hline
 \end {array}\right ]
 \,
   \left [\begin {array}{c}
  \hline
    U_{-{\cal N}-1}
 +
 {g}^2U_{-{\cal N}-3}
 \\
 (1-{g}^2)\psi_{-{\cal N}}\\
 \hline
 \end {array}\right ]=
  \left [\begin {array}{c}
  \hline
   (1-{g}^2)U_{-{\cal N}-2}\\
   (1-{g}^2)\psi_{-{\cal N}+1}\\
  \hline
 \end {array}\right ]
 \,,
 \een
 \ben
 \left [\begin {array}{cc} \hline
    2\cos \varphi&-1
 \\
 -1&2\cos \varphi
 \\
 \hline
 \end {array}\right ]\,
   \left [\begin {array}{c}
  \hline
  (1-{g}^2) {\psi}_{\cal N}
 \\
  L_{{\cal N}+1}+{g}^2L_{{\cal N}+3} \\
  \hline
 \end {array}\right ]=
  \left [\begin {array}{c}
  \hline
  (1-{g}^2){\psi}_{{\cal N}-1}\\
  (1-{g}^2)L_{{\cal N}+2}\\
  \hline
 \end {array}\right ]
 \,
 \een
which can be simplified to read
 \ben
 (1-{g}^2)\,\psi_{-{\cal N}}=
 U_{-{\cal N}}+2{g}^2U_{-{\cal N}-2}+{g}^2U_{-{\cal N}-4}\,,
 \een
 \ben
 (1-{g}^2)\,\psi_{-{\cal N}-1}=
 U_{-{\cal N}-1}+{g}^2U_{-{\cal N}-3}\,,
 \een
 \ben
 (1-{g}^2)\,\psi_{{\cal N}}=
 L_{{\cal N}}+2{g}^2L_{{\cal N}+2}+{g}^2L_{{\cal N}+4}\,,
 \een
 \ben
 (1-{g}^2)\,\psi_{{\cal N}+1}=
 L_{{\cal N}+1}+{g}^2L_{{\cal N}+3}\,.
 \een
These equations represent the two alternative definitions of the
sum $C+D$ and of the difference $C-D$ of the two unknown
coefficients in $\psi_k$,
 \ben
 2\,(1-{g}^2)(C+D)\cos {\cal N} \varphi=A^*(\varphi)+A(\varphi)\,(R+T)\,,
 \een
 \ben
 2\,(1-{g}^2)(C+D)\cos ({\cal N}+1) \varphi=B^*(\varphi)+B(\varphi)\,(R+T)\,
 \een
 \ben
 -2\,{\rm i}\,(1-{g}^2)(C-D)\sin {\cal N} \varphi=A^*(\varphi)+A(\varphi)(R-T)\,,
 \een
 \ben
 -2\,{\rm i}\,(1-{g}^2)(C-D)\sin ({\cal N}+1) \varphi=B^*(\varphi)+B(\varphi)\,(R-T)\,
 \een
where we abbreviated
 \ben
 A(\varphi)=e^{{\rm i}{\cal N}\,\varphi}+{g}^2
 \left (2\,e^{{\rm i}({\cal N}+2)\,\varphi}+e^{{\rm i}({\cal N}+4)\,\varphi}
 \right)\,,
 \een
 \ben
 B(\varphi)=e^{{\rm i}({\cal N}+1)\,\varphi}+{g}^2
 \,e^{{\rm i}({\cal N}+3)\,\varphi}
 \,.
 \een
In the next step we eliminate $C$ and $D$ and express
 \ben
 R-T=-\frac{u^*(\varphi)}{u(\varphi)}\,,
 \ \ \ \ \ \ \
 u(\varphi)=\frac{B(\varphi)}{\sin ({\cal N}+1) \varphi}-
 \frac{A(\varphi)}{\sin {\cal N} \varphi}\,,
 \een
 \ben
 R+T=-\frac{v^*(\varphi)}{v(\varphi)}\,,
 \ \ \ \ \ \ \
 v(\varphi)=\frac{B(\varphi)}{\cos ({\cal N}+1) \varphi}-
 \frac{A(\varphi)}{\cos {\cal N} \varphi}\,.
 \een
The required amplitudes $R$ and $T$ are now found, in closed form,
as the respective sum and difference of the latter two
expressions.  In the final step their probability conservation
property
 \ben
 |R|^2+|T|^2=1
 \een
is easily seen.

\section{Summary \label{summary}}

Our main technical result is that via a discretization of the real
axis of coordinates $x$ (and using the matching method) an exact
linear-algebraic solvability of our present model of scattering has
been achieved. Constructively, the necessary unitarity requirement
has been satisfied at the same time. Our model [containing several
spatially separated and strictly localized interactions which {\em
appear} non-Hermitian in ${\cal H}^{(F)}\, \equiv \,L_2(I\!\!R)$] is
being assigned the more or less unique Hilbert space of states
${\cal H}^{(S)}\,\equiv\,{\cal H}^{(physical)}$ where the use of an
anomalous inner product makes the Hamiltonians (crypto){Hermitian}.

It is worth noticing that the metric operator which defines the
inner product in ${\cal H}^{(physical)}$ merely differs from the
usual Dirac's delta function {\em locally}, viz., in a close
vicinity of interaction points. This implies that the physical
operator of the coordinate remains unmodified almost everywhere. An
entirely consistent physical picture of scattering from multiple
scatterers is obtained in this way. In contrast to the older models
using non-Hermitian but strictly local potentials $V(x)$, the {free
motion} between our present, slightly nonlocal individual
non-Hermitian point-like scatterers remains undistorted.

In conclusion let us re-emphasize that the motivation and
inspiration of our present study of a simplified model of multiple
scattering resulted from several sources. One of the most important
ones has to be seen in the recent enormous growth of interest in the
models of quantum dynamics of bound states which look {\em
manifestly non-Hermitian} in $L_2(I\!\!R)$ and/or in similar
mathematical representations of the Hilbert space of states
\cite{Carl}.

The key to success can be seen in the discovery of feasibility of a
strictly physics-motivated transition to correct Hilbert space
${\cal H}^{(physical)}$ \cite{SIGMA}. Our present paper can be read
as an implementation and advertisement of such an approach where one
chooses a slightly more complicated input physics (i.e., in our
case, a slightly nonlocal Hamiltonian $H=T+V$) and where one is
rewarded by a perceivable simplification of mathematics. In
particular, we saw that the resulting metric $\Theta$ in ${\cal
H}^{(physical)}$ differed from the unit operator just in a finite
number of matrix elements in our model.

We have only to repeat that our second, less abstract motivation
grew from the emergence of several very recent studies of manifestly
non-Hermitian models of quantum scattering \cite{Cannata}. Several
issues may be addressed in this context. For example, in the less
ambitious, effective-theory versions of these models (where one does
not insist on the conserved probability) one can easily stay in the
single, effective-theory Hilbert space ${\cal H}^{(F)}$. Moreover,
various additional dynamical assumptions (like the strict locality
of potentials $V(x)$) may easily be incorporated in the similar
pragmatic applications of the theory.

In contrast, in the ``fundamental" and unitary quantum theory a real
challenge is to be seen in the existence of a correlation between
non-Hermiticity of a local $V$ and the long-range non-locality
emerging in $\Theta$ \cite{Mostafazadeh}. One can notice that this
relationship seems highly model-dependent. In this sense, our
present message can be read as a methodical encouragement.
Basically, we found that whenever one broadens the class of the
eligible potentials the latter model-dependence can be
re-interpreted as an advantage.

It should be remembered that the increase of the non-Hermiticity of
$H$ need not necessarily be correlated with the growth of
non-localities in $\Theta$ obscuring the clear physical picture of
scattering. We succeeded here in showing that {\em both} the
non-localities occurring in $V$ and $\Theta$ can be  kept under
control {\em simultaneously}. After all, one may note that in our
present one-dimensional model with $x \in I\!\!R$ the anomalies
disappear ``almost everywhere" in the continuous limit $h \to 0$.

In this manner our present text brought a rather surprising
resolution of the puzzle formulated in ref.~\cite{FT} where we did
not manage to get rid of the non-locality in a non-Hermitian model
comprising several {spatially separate} scattering centers. Here we
revealed that sometimes it makes good sense to {\em sacrifice} some
inessential symmetries of the model in order to preserve either its
exact solvability or its phenomenological flexibility. We should
note that the feasibility of our (computer-assisted) algebraic
manipulations survived even the transition to unusually complicated
point-interaction simulations by three-by-three matrices.

In the context of physics good news concern, first of all, the
possibility of an explicit construction of an optimal metric
$\Theta$ in the physical Hilbert space. Its ``optimality" reflects
the fact that with an obvious exception of the closest vicinities of
the point-like interaction centers of our model, the metric $\Theta$
itself has successfully been {forced to commute} with the operator
of the coordinate almost everywhere. This means that in contrast to
intuitive expectations (supported even by some solvable models), the
concept of coordinate and of an asymptotically free (i.e.,
measurable) motion of a quantized object can survive the emergence
of a finite number of point-like non-Hermitian obstacles positioned
arbitrarily along the real line.

The latter observation allows us to declare that our model
represents an illustrative example of a standard quantum system
where the non-Hermiticity as well as the resulting non-localities
(in both the metric $\Theta$ and in wave functions) remain confined
to a very small part of the domain of the coordinates. This means
not only that up to the singular points the coordinates remain
measurable but also that the clear physical picture and consistent
probabilistic interpretation of the non-Hermitian systems is
naturally being extended to the multiple-scattering scenario.

 \vspace{5mm}

\section*{Acknowledgement}

Work supported  by the Institutional Research Plan AV0Z10480505, by
the GA\v{C}R grant Nr. 202/07/1307 and  by the M\v{S}MT ``Doppler
Institute" project Nr. LC06002.

\section*{Appendix A. Construction of the amplitudes for model  (\ref{propo})
at ${\cal N}=-1$ }

In the special case of our toy model $H^{(g,{\cal N})}$ at ${\cal
N}=-1$ the analysis of the respective transition and reflection
amplitudes $T$ and $R$ can be based on the explicit solution of
Schr\"{o}diner equation which degenerates, in an obvious manner and
under the notation conventions of section \ref{secIV}, to the
following set of the five linear relations representing matching
conditions near the origin,
 \ben
\left [\begin {array}{rcccccr} \hline
    -1&2\cos \varphi&-1-{g}&0&0&0&0
 \\
 0&-1+{g}&2\cos \varphi&-1+{g}&0&0&0
  \\
 0&0&  -1-{g}&2\cos \varphi&-1-{g}&0&0
 \\
 0&0&0&-1+{g}&2\cos \varphi&-1+{g}&0
 \\
 0&0&0&0&-1-{g}&2\cos \varphi&-1\\
 \hline
 \end {array}\right ]\,
  \left [\begin {array}{c}
  \hline
  U_{-3}\\
  U_{-2}
 \\
 \hline
  U_{-1}
 +
 \chi_{-1}
 \\
 \psi_0
 \\
  L_1+\chi_1 \\
  \hline
  L_2 \\
  L_3\\
  \hline
 \end {array}\right ]=0\,.
 \,
 \een
Their solution may start from the  first and last line giving
 \ben
 (1+{g})\,\chi_{-1}=-{g}\,U_{-1}=-{g}\,(e^{-{\rm i}\varphi}
 +R\,e^{{\rm i}\varphi})\,,\ \
 \ \ \ \ \
 (1+{g})\,\chi_{1}=-{g}\,L_1=-{g}\,T\,e^{{\rm i}\varphi}\,.
 \een
This enables us to consider just the three modified matching
conditions
 \ben
\left [\begin {array}{ccccc}
     -1+{g}^2&2\cos \varphi&-1&0&0
  \\
 0&  -1+{g}^2&2\cos \varphi&-1+{g}^2&0
 \\
 0&0&-1&2\cos \varphi&-1+{g}^2
  \end {array}\right ]\,
  \left [\begin {array}{c}
  \hline
  U_{-2}
 \\
 \hline
  U_{-1}
 \\
 (1-{g}^2)\,
 \psi_0
 \\
  L_1 \\
  \hline
  L_2 \\
  \hline
 \end {array}\right ]=0
 \,.
 \een
The first and last rows read
 \ben
 (1-{g}^2)\psi_0
 =U_0+{g}^2U_{-2}\,=
 1+{g}^2\, e^{-2{\rm i}\varphi}
 +
 (1+{g}^2\,e^{2{\rm i}\varphi })\,R\,
 \een
 \ben
 (1-{g}^2)\psi_0
 =L_0+{g}^2\,L_2=
 (1+{g}^2\,e^{2{\rm i}\varphi })\,T\,
 \een
so that their combination
 \ben
  1+{g}^2\, e^{-2{\rm i}\varphi}
 =
 (1+{g}^2\,e^{2{\rm i}\varphi })\,(T-R)\,
 \een
defines the difference between our two amplitudes as a complex
number with  unit norm,
 \ben
 T-R=\frac{1-{\rm i}\lambda}{1+{\rm i}\lambda}\,\equiv\,e^{{\rm i}\alpha}\,,
 \ \ \ \ \lambda=\frac{{g}^2\sin 2\varphi}{1+{g}^2\cos 2\varphi}
 \,.
 \een
The remaining central matching condition can be given the form of an
equation for the sum $\Sigma=T+R$ of the amplitudes, with the
solution equal to another complex number with unit norm,
 \ben
 T+R=-e^{-2{\rm i}\varphi}\,\frac{1-{\rm i}\mu}{1+{\rm i}\mu}
 \,\equiv\,e^{{\rm i}\beta}\,,
 \ \ \ \ \mu=\frac{(1-{g}^2)\,\sin 2\varphi}{1-3{g}^2-\cos 2\varphi
 -{g}^2\cos 2\varphi}
 \,.
 \een
This gives the two final formulae
 \ben
 2\,T=e^{{\rm i}\beta}+e^{{\rm i}\alpha}\,,\ \ \ \ \ \ \ \
 2\,R=e^{{\rm i}\beta}-e^{{\rm i}\alpha}\,
 \een
with the two respective properties
 \ben
 4\,|T|^2=
 \left (e^{{\rm i}\beta}+e^{{\rm i}\alpha}
 \right )\,
 \left (e^{-{\rm i}\beta}+e^{-{\rm i}\alpha}
 \right )=2+e^{{\rm i}(\alpha-\beta)}+e^{{\rm i}(\beta-\alpha)}
 \een
 \ben
 4\,|R|^2=
 \left (e^{{\rm i}\beta}-e^{{\rm i}\alpha}
 \right )\,
 \left (e^{-{\rm i}\beta}-e^{-{\rm i}\alpha}
 \right )=2-e^{{\rm i}(\alpha-\beta)}-e^{{\rm i}(\beta-\alpha)}
 \een
which imply that
 \ben
 |R|^2+|T|^2 =1\,.
 \een
This means that in contrast to the observations made in some other
non-Hermitian models \cite{Jonesdva,Jones,prd}, the flow of
probability is conserved so that the standard physical picture of
the scattering does not require any modifications.

\section*{Appendix B. Construction of the amplitudes for model  (\ref{propo})
%
at  ${\cal N}=0$}


In place of the five-dimensional matching condition of our preceding
appendix let us now turn our attention to the family of nontrivial
models where the two three-dimensional elementary-interaction
submatrices are separated by a free-motion interval of the length
$2{\cal N}+1$. In the first nontrivial model with ${\cal N}=0$ the
nonvanishing submatrix of our interaction matrix is
seven-dimensional,
 \ben
 V^{(g,\,0)}=
\left [\begin {array}{rrr|c|rrr} \hline
 0&g&0&{\mbox{\,}}0{\mbox{\,}}&0&0&0
  \\ \!-g&0&\!-g&0&0&0&0
 \\
 0&g&0&0&0&0&0
  \\
  \hline
 0&0&  0&0&0&0&0
 \\
 \hline
 0&0&0&0&0&g&0
 \\
 0&0&0&0&\!-g&0&\!-g\\
 0&0&0&0&0&g&0
 \\
 \hline
 \end {array}\right ]\,.
 \een
In such a case one has to consider seven matching conditions of the
form
 \ben
 M^{[0]}( \varphi)\,
   \left [\begin {array}{c}
  \hline
  U_{-3}\\
  U_{-2}
  +
 \chi_{-2} \\
 \hline
  U_{-1}
 +
 \chi_{-1}
 \\
 \psi_0
 \\
  L_1+\chi_1 \\
  \hline
  L_2+
 \chi_{2} \\
  L_3\\
  \hline
 \end {array}\right ]=
  \left [\begin {array}{c}
  \hline
  U_{-4}\\
  0\\
  \hline 0\\ 0\\ 0\\
  \hline 0\\
  L_4\\
  \hline
 \end {array}\right ]
 \,
 \een
where
 \ben
 M^{[0]}( \varphi)=
\left [\begin {array}{ccc|c|ccc} \hline
    2\cos \varphi&-1-{g}&0&0&0&0&0
 \\
    -1+{g}&2\cos \varphi&-1+{g}&0&0&0&0
 \\
 0&-1-{g}&2\cos \varphi&-1&0&0&0
  \\ \hline
 0&0&  -1&2\cos \varphi&-1&0&0
 \\
  \hline
 0&0&0&-1&2\cos \varphi&-1-{g}&0
 \\
 0&0&0&0&-1+{g}&2\cos \varphi&-1+{g}\\
 0&0&0&0&0&-1-{g}&2\cos \varphi\\
 \hline
 \end {array}\right ]
 \,.
 \een
The separate subset of the  first and last matching condition is
solvable as follows,
 \ben
 (1+{g})\,\chi_{-2}=-{g}\,U_{-2}\,,\ \
 \ \ \ \ \
 (1+{g})\,\chi_{2}=-{g}\,L_2\,.
 \een
The backward insertion of these formulae leads to the quintuplet of
the reduced matching conditions
 \ben
\left [\begin {array}{ccccccc} \hline
    -1+{g}^2&2\cos \varphi&-1+{g}^2&0&0&0&0
 \\
 0&-1&2\cos \varphi&-1&0&0&0
  \\
 0&0&  -1&2\cos \varphi&-1&0&0
 \\
 0&0&0&-1&2\cos \varphi&-1&0
 \\
 0&0&0&0&-1+{g}^2&2\cos \varphi&-1+{g}^2\\
 \hline
 \end {array}\right ]
   \left [\begin {array}{c}
  \hline
  U_{-3}\\
  U_{-2}
   \\
 \hline
  U_{-1}
 +
 \chi_{-1}
 \\
 \psi_0
 \\
  L_1+\chi_1 \\
  \hline
  L_2 \\
  L_3\\
  \hline
 \end {array}\right ]=0
 \,.
 \een
Its first and last line define the other two correction components,
 \ben
 (1-{g}^2)\,\chi_{-1}={g}^2
 \left (
 U_{-1}+U_{-3}
 \right )\,,\ \
 \ \ \ \ \
 (1-{g}^2)\,\chi_{1}={g}^2
 \left (
 L_{1}+L_{-3}
 \right )\,
 \een
so that we are left with the three matching conditions
 \ben
\left [\begin {array}{ccccc} \hline
 -1+{g}^2&2\cos \varphi&-1&0&0
  \\
 0&  -1&2\cos \varphi&-1&0
 \\
 0&0&-1&2\cos \varphi&-1+{g}^2
 \\
 \hline
 \end {array}\right ]
   \left [\begin {array}{c}
  \hline
  U_{-2}
   \\
 \hline
  U_{-1}
 +
 {g}^2 U_{-3}
 \\
 (1-{g}^2)\psi_0
 \\
  L_1+{g}^2L_3 \\
  \hline
  L_2 \\
  \hline
 \end {array}\right ]=0
 \,.
 \een
Their first and last item define the same quantity in two ways,
 \ben
 (1-{g}^2)\,\psi_{0}=U_0+{g}^2
 \left (
 U_{-2}+2\cos \varphi\,U_{-3}
 \right )=
 U_0+{g}^2
 \left (
 2\,U_{-2}+U_{-4}
 \right )
 \,
 \een
 \ben
 (1-{g}^2)\,\psi_{0}=L_0+{g}^2
 \left (
 L_{2}+2\cos \varphi\,L_{3}
 \right )=L_0+{g}^2
 \left (
 2\,L_{2}+L_{4}
 \right )\,.
 \een
In effect, one can eliminate $\psi_0$,
 \ben
 (T-R)\, \left [
 1+{g}^2
 \left (
 2\,e^{2{\rm i}\varphi}+e^{4{\rm i}\varphi}\,
 \right )
 \right ]= \left [
 1+{g}^2
 \left (
 2\,e^{-2{\rm i}\varphi}+e^{-4{\rm i}\varphi}\,
 \right )
 \right ]
 \,
 \een
and specify the difference between $T$ and $R$,
 \ben
 T-R=\frac{1-{\rm i}\lambda'}{1+{\rm i}\lambda'}\,\equiv\,e^{{\rm i}\alpha'}\,,
 \ \ \ \ \lambda'=\frac{{g}^2(2\sin 2\varphi+\sin 4\varphi)}
 {1+{g}^2(2\cos 2\varphi+\cos 4\varphi)}
 \,.
 \een
Next, in a complete parallel to the previous construction, the sum
$\Sigma$ of $T$ and $R$ may and should be extracted again from the
last and symmetrized middle item of our matching conditions,
 \ben
 2\,U_{-1}+2\,L_1
 +2\,{g}^2
 \left (U_{-3}+L_3\right )
 =U_0+L_0+
 {g}^2\,
 \left (2\,U_{-2}+2\,L_2+
 U_{-4}+L_4
 \right )
 \,.
 \een
After appropriate insertions this gives the similar formula as
above,
 \ben
 T+R=-\frac{1-{\rm i}\mu'}{1+{\rm i}\mu'}
 \,\equiv\,e^{{\rm i}\beta'}\,,
 \ \ \ \ \mu'=\frac{-2\sin \varphi+
 {g}^2(2\sin 2\varphi-2\sin 3\varphi+\sin 4\varphi
 )}{[1-2\cos \varphi+
 {g}^2(2\cos 2\varphi-2\cos 3\varphi+\cos 4\varphi
 )
 ]}
 \,.
 \een
The same argumentation as above confirms the validity of the
identity
 \ben
 |R|^2+|T|^2 =1\,
 \een
i.e., of the same probability conservation law as above.


\newpage

\begin{thebibliography}{99}

\bibitem{Jonesdva}
H. F. Jones, 
Phys. Rev. D 78, 065032 (2008).

%

\bibitem{Jones}
H. F. Jones, Phys. Rev. D 76, 125003 (2007).

\bibitem{cubic}
A. Mostafazadeh, J. Phys. A: Math. Gen. 39, 10171  (2006).

\bibitem{prd}
M. Znojil, 
Phys. Rev. D 78, 025026 (2008).

\bibitem{RK}
F. S. Acton, Numerical Methods that Work (Harper \& Row, New York,
1970);

%
M. Znojil, Phys. Lett. A 223, 411 (1996);

M. F. Fern´andez, R. Guardiola, J. Ros and M. Znojil, J. Phys. A:
Math. Gen. 32, 3105  (1999);

M. Znojil,
J. Phys. A: Math. Gen. 39, 10247 (2006).


\bibitem{Christo}
C. Figueira de Morisson Faria and A. Fring,
Laser Physics 17, 424  (2007);

M. Berry, J. Phys. A: Math. Theor. 41, 244007 (2008);

D. Krej\v{c}i\v{r}\'{\i}k and M. Tater, J. Phys. A: Math. Theor. 41,
244013 (2008);

Z. H. Musslimani, K. G. Makris, R. El-Ganainy and D. N.
Christodoulides, Phys. Rev. Lett. 100, 030402 (2008) and J. Phys. A:
Math. Theor. 41, 244019 (2008).

\bibitem{Siegl}
C. M. Bender and K. A. Milton, 
Phys.Rev. D 57, 3595  (1998);

M. Znojil,
J. Phys. A: Math. Gen. 35, 2341 (2002) and
%
Nucl. Phys. B 662, 554 (2003);

P. Siegl, J. Phys. A: Math. Theor. 41, 244025 (2008).

\bibitem{Cannata}
Z. Ahmed, Phys. Lett. A 324, 152 (2004);

Z. Ahmed, C. M. Bender and M. V. Berry,
 J. Phys. A: Math. Gen. 38, L627  (2005);

M. Znojil,
J. Phys. A: Math. Gen. 39, 13325 (2006);

%
F. Cannata, J.-P. Dedonder and A. Ventura, Ann. Phys. 322, 397
(2007);

A. Lavagno, J. Phys. A: Math. Theor. 41, 244014  (2008);

G. L\'evai, P. Siegl and M. Znojil, 
J. Phys. A: Math. Theor. 42 (2009), in print
(arXiv:0906.2092).


\bibitem{Rotter}
H. B. Geyer and I. Snyman, Czech. J. Phys. 55, 1091 (2005);

F. G. Scholtz and H. B. Geyer, J. Phys. A: Math. Gen. 39, 10189
(2006);

P. E. G. Assis and A. Fring, J. Phys. A: Math. Theor. 41, 244001
(2008);

I. Rotter, J. Phys. A: Math. Theor 42, 153001 (2009).

\bibitem{Kamen}
A.Mostafazadeh, Czech. J. Phys. 54, 93 (2003);

A. A. Andrianov, F. Cannata and A. Y.Kamenshchik, J. Phys. A: Math.
Gen. 39, 9975 (2006).


\bibitem{Geyer}
F. G. Scholtz, H. B. Geyer and F. J. W. Hahne, Ann. Phys. (NY) 213,
 74 (1992).


\bibitem{SIGMA}
M. Znojil,
SIGMA 5, 001 (2009).

\bibitem{Ali}
A. Mostafazadeh, J. Math. Phys. 43, 205  and 2814 (2002);

A. Mostafazadeh and A. Batal, J. Phys. A: Math. Gen. 37, 11645
(2004);

U. Guenther, F. Stefani and M. Znojil,
J. Math. Phys. 46, 063504 (2005);

V. Jakubsk\'{y} and J. Smejkal, Czech. J. Phys. 56, 985 (2006).

\bibitem{Smilga}
A. V. Smilga, 
J. Phys. A: Math. Theor. 41, 244026  (2008).

\bibitem{FT}
M. Znojil, 
J. Phys. A: Math. Theor. 41, 292002  (2008).

\bibitem{Carl}
P. Dorey, C. Dunning and R. Tateo,
%
J. Phys. A:Math. Gen. 40, R205 (2007) (arXiv:hep-th/0703066) and
Pramana, to appear 
 (arXiv:0906.1130);

%
C. M. Bender, Rep. Prog. Phys. 70, 947 (2007)
(arXiv:hep-th/0703096);

A. Mostafazadeh, Pseudo-Hermitian Quantum Mechanics,
%
arXiv:0810.5643.
%

\bibitem{Mostafazadeh}
A. Mostafazadeh, J. Phys. A: Math. Gen. 39, 13495  (2006) plus
private communication (2008).


\end{thebibliography}
\end{document}